\documentclass{article}

\usepackage[margin=1.2in]{geometry}
\usepackage[utf8]{inputenc} % allow utf-8 input
\usepackage[T1]{fontenc}    % use 8-bit T1 fonts
\usepackage{hyperref}       % hyperlinks
\usepackage{url}            % simple URL typesetting
\usepackage{booktabs}       % professional-quality tables
\usepackage{enumitem}
\usepackage{amsmath,amsfonts,amsthm,amssymb}
\usepackage{leftindex}
\usepackage{centernot}
\usepackage{diagbox}

\usepackage{nicefrac}       % compact symbols for 1/2, etc.
\usepackage{microtype}      % microtypography
\usepackage{mathabx}
\usepackage{mathtools}
\usepackage{xfrac}
\usepackage{csquotes}
\usepackage{algorithm}
\usepackage{algpseudocode}

\usepackage{bbm}

\usepackage[]{xcolor}
\usepackage[numbers]{natbib}

\newcommand{\EE}{\mathbb{E}}
\newcommand{\bbP}{\mathbb{P}}

\newtheorem{remark}{Remark}
\newtheorem{example}{Example}

\usepackage{graphicx} 
\usepackage{caption}
\usepackage{subcaption}
\usepackage{enumitem}
\usepackage{cleveref}
\usepackage{numprint}
\usepackage{multirow}
\usepackage{makecell}

\title{Anytime-valid testing with e-values and\\ confirmatory adaptive designs}

\author{%
  Werner Brannath and Lasse Fischer\\
   University of Bremen \\
  \texttt{brannath@uni-bremen.de, fischer1@uni-bremen.de}
}

\begin{document}

\maketitle
\begin{abstract}
Confirmatory adaptive designs were introduced more than 30 years ago and enable for example sample size re-assessments and the selection of treatments, endpoints as well as subpopulations during the course of a clinical trial. Recently, sequential tests based on e-values for an anytime-valid inference have been developed, promising seemingly similar or even more flexibility and utility. In this note, we compare these two independently developed concepts, shedding light on their formal and methodological connections and differences. Specifically, we show that adaptive design tools like conditional error functions and combination tests are formally equivalent to e-value based, anytime-valid sequential tests. However, in spite of their common fundamental intention to bring flexibility into statistical inference, they have quite different emphases: While hypothesis testing with combination tests and conditional error function usually intent to exhaust type I error rates under the offered flexibility, e-value based testing aims on the additional flexibility with regard to optional continuation, the chosen level and, in recent extensions, in the loss functions to be controlled. 
% Tun wir das folgende wirklich?
We also indicate how recent e-value achievements could enrich clinical trial methodology and adaptive design methodology could inspire and improve e-value based testing.
\end{abstract}

% \tableofcontents
% \newpage

\section{Introduction}

The powerful e-value theory for anytime-valid inference, which emerged in the recent years \citep{grunwald2020safe, ramdas2023game, shafer2021testing, vovk2021values}, permits to stop and test 
(or continue) with accumulating data at any time during the data accumulation process. An important characteristics is the fact that, in order to control the probability for false rejections, the data driven stopping time does not need to be prespecified. The flexibility with regard to the sample size reminds one to the flexibility of confirmatory adaptive designs (in the following, we just write adaptive designs for brevity) \citep{bauer1989multistage,proschan1995designed,wassmer2016group}, in which (among others) data-driven adaptations of sample size and number of stages without a prespecification of the adaption rule are permitted as well. Adaptive designs are primarily developed for clinical trials, in particular for phase II and phase III clinical trials and combinations thereof. With the conditional rejection probability principle \cite{muller2001comb,muller2004general} and recursive combination tests \cite{brannath2002recot} such adaptions can be done at any time during the data accumulation process. This leads to the natural question of how these seemingly different approaches are related. This question is addressed in this note.

We will show that concepts of adaptive designs, like combination tests and conditional error functions, are formally equivalent to anytime-valid test based on sequential e-values. We also illustrate a fundamental difference of the two approaches in their emphasis: While combination tests and conditional error functions intent to exhaust type I error rate levels, meaning to have a type I error probability of exactly $\alpha$, e-values intend to introduce flexibility with regard to optional continuation, choice of type I error rate levels and, more general, the expected losses to be controlled. These two emphases are complementary in the sense that the first cannot be achieved without violating the second and vice versa. Moreover, the e-value based perspective has the potential to provide solutions where the adaptive design approach may fail because the latter requires more information on the null distribution of the test statistics than required for e-value based tests.  

It should be noted that we do not claim to derive substantially new methodology in this paper but show the yet unexplored connections and differences between the closely related approaches of e-values and adaptive designs. By pointing out connections, we aim to build a bridge between the yet rather weakly connected scientific communities and literature.  In addition, recent achievements of e-values are highlighted in the discussion that may also be valuable to the adaptive design community. On the other hand, achievements in the field of adaptive designs may inspire new achievements and improvements for e-value based sequential tests, like the uniform improvement via recursive combinations tests presented in the Appendix.

\paragraph{Anytime-valid inference via e-values.}
We start with a brief and basic introduction to anytime-valid inference via the e-value methodology. Let us assume that we are interested in testing a null hypothesis $H_0$ with a possibly infinite stream of data. From this stream of data we calculate over time a sequence of non-negative statistics $E_1, E_2, \ldots$, where each $E_t$ can either be calculated based on one data point or a batch of data, with the property that at any time point $t$
\begin{align}\label{eq:evprop}
\EE_{\mathbb{P}}\left[E_t|E_1,\ldots, E_{t-1}\right]\leq 1 \text{ for all distributions $\mathbb{P}\in H_0$} .
\end{align}
This particularly allows to choose each $E_t$ based on the data (and external information) of the stages $1,\ldots, t-1$, as long as \eqref{eq:evprop} is satisfied.
A sequence of statistics satisfying \eqref{eq:evprop} is often denoted as ``sequential e-values'' \citep{vovk2021testing}, whereby an ``e-value'' is understood to just be a nonnegative random variable with expected value at most $1$ under $H_0$. Sequential e-values can be used for anytime-valid inference in the following way: Defining the test statistics $M_0=1$,
\begin{align}\label{eq:eproc}
M_t=\prod_{i=1}^t E_i , \quad t=1,2,\ldots ,
\end{align}
we control the type I error at level $\alpha$ when rejecting $H_0$ if $M_t\geq 1/\alpha$ at any time $t$. This follows from Ville's inequality \citep{ville1939etude} (a sequential generalization of Markov's inequality), since $(M_t)_{t\in \mathbb{N}_0}$ is a nonnegative supermartingale under $H_0$  with $M_0= 1$.  Such a process $(M_t)_{t\in \mathbb{N}_0}$ is also called \textit{test supermartingale} \citep{vovk2005algorithmic, shafer2011test}. Furthermore, by the optional stopping theorem $\mathbb{E}_{\mathbb{P}}[M_{\tau}]\leq 1$ for any stopping time $\tau$ and $\mathbb{P}\in H_0$, showing that every stopped test supermartingale is an e-value. 

% Moreover, by Doob's optional stopping theorem, we obtain that $\EE[M_\tau]\leq 1$ for any stopping time $\tau$. A stochastic process $(M_t)_{t\in \mathbb{N}}$ that satisfies the latter property is called an "e-process".

\begin{example}[Likelihood ratios]\label{example:SPRT_1}
    Suppose we have an i.i.d. sequence of data $X_1,X_2,\ldots$ and are testing a simple null hypothesis against a simple alternative. Let $p_{0}$ be the density of the null distribution and $p_{1}$ the alternative density. Then the likelihood ratios 
    $
    \lambda_t=\frac{p_{1}(X_t)}{p_{0}(X_t)}, t\in \mathbb{N},
    $
    define sequential e-values, since $$\mathbb{E}_{\mathbb{P}_0}[\lambda_t|\lambda_1,\ldots,\lambda_{t-1}]=\int_{\text{supp}(p_{0})} \frac{p_{1}(x)}{p_{0}(x)}p_{0}(x)\ \mu(dx)=\int_{\text{supp}(p_{0})} p_{1}(x)\ \mu(dx)\leq 1.$$ 
The resulting anytime-valid test is equivalent to the sequential probability ratio test (SPRT) by \citet{wald1945sequential}, if no lower bound to accept $H_0$ is specified. Likelihood ratios play an important role in the e-value theory, since log-optimal e-values always take the form of a likelihood ratio, even if the null hypothesis is composite \citep{grunwald2020safe, larsson2024numeraire}. In Section~\ref{sec:discussion} we will give a brief description of log-optimality. 

It is valuable to note that $p_0$ in $\lambda_t$ can be the maximum over the densities of a composite null hypothesis $H_0$ without violating the e-value property of $\lambda_t$. Moreover, the numerator $p_1$ could be the maximum of the densities from some composite alternative determined with an independent training sample. These ideas have been introduced in \cite{wasserman2020universal} and are known as ``universal inference''.    
\end{example}

\begin{remark}\label{remark:general_e}
There are sequential tests that cannot be constructed by multiplying sequential e-values in the way described above \citep{ramdas2022testing}. In general, one would need to define a sequential e-value $E_t^{\mathbb{P}}$ for each null distribution $\mathbb{P}\in H_0$, multiply those sequential e-values $M_t^{\mathbb{P}}=\prod_{i=1}^t E_i^{\mathbb{P}}$ and then define $(M_t)_{t\in \mathbb{N}_0}$ such that $M_t\leq M_t^{\mathbb{P}}$ $\mathbb{P}$-almost surely for all $\mathbb{P}\in H_0$ and $t\in \mathbb{N}_0$ \citep{ramdas2020admissible, ramdas2023game}. Although $(M_t)_{t\in \mathbb{N}_0}$ is not necessarily a test supermartingale in this case, it is easy to show that Ville's inequality and the optional stopping theorem still apply, such that $(M_t)_{t\in \mathbb{N}_0}$ can be used in the same way for testing as before. However, in most testing problems that are relevant for clinical trials, powerful sequential tests can be constructed by calculating a single sequential e-value for each stage  \citep{grunwald2020safe, ly2024tutorial, ramdas2024hypothesis}. To reduce notation and  keep it simple, we focus on such constructions in the following.
\end{remark}

% An anytime-valid test allows to look into the data at any time during the course of the trial and make a decision on the null hypothesis based on the data observed so far, while controlling the type I error probability. Sequential e-values (or equivalently the multi-stage conditional error principle described in the previous section) allow for such anytime-valid inference by calculating a sequential e-value $E_t$ (or a conditional error function $A_t$) for each individual data point. A common misconception is that one must pay a price in form of statistical efficiency when allowing for anytime-valid inference. However, this is not true, as \citet{muller2004general} (and more recently \citet{koning2025sequentializing} in the context of e-values) show that every test can be embedded into an anytime-valid test. Indeed, the converse is true, one can gain statistical efficiency by using an anytime-valid test. For example, it is long known \citep{wald1945sequential, wald1948optimum}, that in case of simple hypotheses, Wald's SPRT, which is an anytime-valid test, beats the Neyman-Pearson test, which is the most powerful fixed-sample test, in terms of expected sample size for predefined type I and type II error probabilities. 

\paragraph{Adaptive designs.}
Next, we give a brief introduction to adaptive design methodology whereby we focus in this section on two-stage designs, $t=1,2$, following the so called \enquote{conditional error function principle} \citep{proschan1995designed} that formally covers all other adaptive design approaches \cite{wassmer2016group,poschbauer1999}. This means that we start with adaptive two-stage designs, where after a first stage (with a preplanned design), also called \enquote{interim analysis}, the design of the second stage can be flexibly specified based on the first stage data. An example for such a design choice is second stage sample size. Other choices (like test statistics) are also possible.

In order to preserve the type I error rate, we prespecify a conditional error function $A$, which is a statistic
calculated from the first stage data, with the property that 
\begin{align}\label{eq:conderr1}
0\le A\le 1\text{ for all interim data,\quad and }\EE_\mathbb{P}[A]\leq \alpha\text{ for all distributions } \mathbb{P}\in H_0 \,. 
\end{align}
Further, let $\psi$ (with values in $\{0,1\}$) be an arbitrary test decision function that is based on all data and satisfies
\begin{align}\label{eq:conderr2}
\EE_{\mathbb{P}}[\psi|A]\le A\quad\text{ for all interim data and for all } \mathbb{P}\in H_0 \, . 
\end{align}
When rejecting $H_0$ if $\psi=1$ and accepting it otherwise, properties \eqref{eq:conderr1} and \eqref{eq:conderr2} guarantee that the type I error probability is controlled at level $\alpha$, because they imply that $\mathbb{P}[\psi=1]=\EE_\mathbb{P}[\psi]=\EE_\mathbb{P}\big[\EE_\mathbb{P}[\psi|A]\big]\le \EE_\mathbb{P}[A]\le\alpha$ for all $\mathbb{P}\in H_0$.  
The flexibility of the presented adaptive testing approach stems from the fact that the test decision function $\psi$ can be freely chosen based on the interim data, as long as it satisfies \eqref{eq:conderr2} at all interim points. Note that property \eqref{eq:conderr2} can be satisfied with the sole knowledge of the conditional expectation of $\psi$ given the interim data (or even given only the observed value of $A$). This permits to use some $\psi$ that remains completely unspecified for all interim data except the actually observed ones and hence provides the possibility for unforeseen, data driven design changes.    
We finally note that if $A=1$ for some interim data point, then we can stop and reject $H_0$ already at stage one, since we can choose $\psi=1$. Similarly, if $A=0$, we can stop and accept $H_0$ at stage one, since $\psi=0$ is the only test decision function satisfying \eqref{eq:conderr2}.

\begin{remark}\label{remark:general_A}
    Similar to Remark~\ref{remark:general_e}, in general one could define a different conditional error function $A^{\mathbb{P}}$ and test $\psi^{\mathbb{P}}$ for each $\mathbb{P}\in H_0$, ensuring that $\mathbb{E}_{\mathbb{P}}[\psi^{\mathbb{P}}|A^{\mathbb{P}}]\leq A^{\mathbb{P}}$ and reject $H_0$ if $\psi^{\mathbb{P}}=1$ for all $\mathbb{P}\in H_0$ \citep{muller2004general}.
\end{remark}

Interestingly, even though, as we will show in this paper, conditional error functions and e-values are formally equivalent, likelihood ratios play only a minor role in adaptive designs. The two main approaches to construct a conditional error function are: 
% (1) the conditional error function approach \cite{proschan1995designed}, 
(1) the concept of combination tests \cite{bauer1989multistage,bauer1994adaptive}
and (2) the conditional rejection probability principle \cite{schaefer2001modif,muller2004general}. We briefly recap these with the following two examples.

\begin{example}[Combination tests]\label{example:fisher}
Combination tests are based on p-values. The general concept will be reviewed in  Section~\ref{sec:combination_tests}. A simple example,
which we describe next, is Fisher's product test. 
Assume that the first and second stage data are from independent cohorts, and are summarized by stage-wise p-values, $P_1$ and $P_2$, whereby $P_1$ is calculated from the first and $P_2$ from the second stage data only. 
$H_0$ is rejected at the first stage if $P_1\le \alpha_1$ for some prespecified $\alpha_1<\alpha$, and at the second stage if $P_1\cdot P_2\leq c$ for some prespecified critical value $c$. 
The critical values $\alpha_1<\alpha$ and $c$ are determined in a way that the rejection probability is equal to $\alpha$
for stochastically independent and uniformly distributed p-values. This will guarantee type I error control also for strictly conservative p-values \citep{brannath2012probabilistic}. The product test approach can be phrased in terms of a conditional error function by defining $A=1$ for $P_1\le \alpha_1$ 
and $A=\min(1,c/P_1)$ otherwise. The test decision function $\psi=\mathbf{1}\{P_1\cdot P_2\le c\}$ will then satisfy \eqref{eq:conderr2} for any valid second stage p-value $P_2$. 
%The approach can be generalized to combination tests 
%where the two stage-wise p-values are combined by another function %$C(P_1,P_2)$ than their product as long as this function is non-decreasing in both arguments and increasing in at least one \citep{brannath2002recot, wassmer2016group}.  
\end{example}

\begin{example}[Conditional  rejection probability principle]\label{example:mueller}
    In the conditional rejection probability principle \cite{muller2001comb,muller2004general,proschan1995designed} we start with a level $\alpha$ test decision function $\phi$ of a fixed sample size or group sequential test. Now let $D$ denote the data at the time we want to do an interim analysis (this time can be chosen based on the data observed so far). Using the general notation of Remark~\ref{remark:general_A}, we can calculate a conditional error function by  $A^{\mathbb{P}}=\mathbb{E}_{\mathbb{P}}[\phi|D]$ for each $\mathbb{P}\in H_0$. Since $\mathbb{E}_{\mathbb{P}}[\phi]\leq \alpha$ for all $\mathbb{P}\in H_0$, it immediately follows that $\mathbb{E}_{\mathbb{P}}[A^{\mathbb{P}}]\leq \alpha$ for all $\mathbb{P}\in H_0$. A similar approach was recently investigated by \citet{koning2026anytime} in the anytime-validity framework.
\end{example}

\paragraph{Paper outline.} We first show how all adaptive designs based on conditional error functions can be written by sequential e-values and vice versa (Section~\ref{sec:equivalence}), showing the equivalence of the two approaches. Afterwards, we provide similar connections between sequential e-value based tests and recursive combination tests. 
% We then also show that with known and continuous least favorable distributions, the latter can uniformly improve the first by exhausting the type I error rates  (Section~\ref{sec:combination_tests}). 
In Section~\ref{sec:multiple_testing}, we consider multiple testing with adaptive designs and e-values, focusing on the weighted
Bonferroni based closed tests considered in \cite{hommel2007short-cuts}.
%graphical approach by \citet{bretz2009graphical}. 
Finally, we discuss recent accomplishments of e-values and how they could enrich the clinical trial methodology in Section~\ref{sec:discussion}, but also point out potential advantages of the older adaptive design methodology.

% This connects these two seemingly different approaches and particularly provides additional intuition on e-value based tests. Afterwards, we point out some properties of e-values and describe how they could enrich the existing adaptive design literature (Section~\ref{sec:discussion}). 

% In the discussion (Section~\ref{sec:discussion}), we highlight three aspects in which we see potentials for e-values to enrich the existing adaptive design literature, namely with regard to their interpretation, topics on multiple testing and aspects like log-optimality.

\section{Equivalence of adaptive designs and e-values based testing\label{sec:equivalence}}

\subsection{Two-stage adaptive designs and e-values\label{sec:two-stage}}

We next show that a two-stage adaptive designs with conditional error function $A$ can be understood in terms of two sequential e-values, namely 
\begin{align}\label{eq:e1e2A}
E_1=\frac{A}{\alpha}\quad \text{ and }\quad  E_2=\frac{\psi}{A}\ .
\end{align}
Obviously, \eqref{eq:conderr1} implies that $\mathbb{E}_\mathbb{P}[E_1]\leq 1$ and \eqref{eq:conderr2} implies that $\mathbb{E}_\mathbb{P}[E_2|E_1]= \mathbb{E}[\psi|A]/A\leq 1$ for all  $\mathbb{P}\in H_0$, showing that $E_1$ and $E_2$ are sequential e-values. Furthermore, if the adaptive design rejects $H_0$ ($\psi=1$), we have $E_1E_2=1/\alpha$, and if the adaptive design accepts $H_0$ ($\psi=0$), we have $E_1E_2=0$, proving that the e-value based test rejects $H_0$ if and only if the adaptive design rejects~$H_0$.

On the other hand, for sequential e-values $E_1$ and $E_2$, we obtain that 
$A=\alpha E_1$ and $\psi=\alpha E_1 E_2$ satisfy \eqref{eq:conderr1} and \eqref{eq:conderr2}. Depending on $E_1$ and $E_2$ it may happen that $A>1$ and $\psi$ takes other values than 0 and 1. However, $A>1$ can be avoided by choosing an $E_1$ that is bounded by $1/\alpha$. Moreover, if we choose $E_2$ to be bounded by $1/(\alpha E_1)$ then $\psi\in [0,1]$ can be interpreted as a decision function of a randomized test \citep{koning2024continuous}. Recall, that we can choose $E_2$ based on the interim data. This test becomes non-randomized if the values of $E_2$ are restricted to $0$ and $1/(\alpha E_1)$. 

This shows an important difference between the emphases of adaptive designs and e-values. Adaptive designs are constructed to exhaust the type I error as much as possible. While e-values allow to exhaust the type I error by solving \eqref{eq:evprop} with an equality and choosing the e-values such that $E_1 E_2\in \{0,1/\alpha\}$, the e-value based tests are often not constructed this way (see also Example~\ref{example:SPRT_1}). There are several reasons why allowing $E_1E_2$ to take other values than $0$ and $1/\alpha$ can be beneficial:
\begin{enumerate}
    \item E-values permit optional continuation \citep{grunwald2020safe, shafer2021testing}. That means if we report $E_1E_2$ even though the product is smaller than $1/\alpha$, one could always decide to continue the trial by multiplying further e-values at a later time. 
    % Even if $E_1E_2$ is smaller than $1/\alpha$, it can be useful to report this product, since $E_1E_2$ is an e-values and large e-values provide evidence against the null hypothesis \citep{shafer2021testing, grunwald2020safe, vovk2021values}. Furthermore, e-values allow for optional continuation. That means, based on those result, one could continue the study at a later time by multiplying further e-values. This also provides a good .Due to the optional stopping theorem, one could later continue the trial   
    \item E-values allow to choose the significance level $\alpha$ or some loss function data-dependently while maintaining a reasonable error control \citep{grunwald2024beyond, koning2023post}. Hence, if $E_1E_2<1/\alpha$, we might still be able to reject at a larger significance level, and if $E_1E_2>1/\alpha$, one could decrease the level $\alpha$ to strengthen the results. The adaptive choice of $\alpha$ can formally be justified by understanding type I error rate control as average control of the loss $L_\alpha:=(1/\alpha)\mathbf{1}\{E_1 E_2\ge 1/\alpha\}$, i.e.\  $\EE_{\mathbb{P}}[L_\alpha]\le 1$ for all $\mathbb{P}\in H_0$, and observing that
    $\EE_{\mathbb{P}}\big[\sup_{\alpha\in (0,1)} L_\alpha\big]=\EE_{\mathbb{P}}[E_1 E_2]\le 1$ for all $\mathbb{P}\in H_0$.      
    \item One branch of the e-value literature particularly focuses on complex settings with composite null hypotheses in which no least favorable configuration exists \citep{wasserman2020universal, grunwald2020safe, larsson2024numeraire}. In those cases it might not be possible to exhaust the type I error probability (with e-values nor adaptive designs) and thus to define e-values with the aforementioned restrictions.
\end{enumerate}
We elaborate on that in the discussion.

In summary, we have shown that every two stage adaptive design can be formulated in terms of sequential e-values, and two stage designs based on suitably defined sequential e-values can be understood as an adaptive design following the conditional error function principle. In the next section we show a similar relationship for multi-stage adaptive designs. 

\begin{remark}
    Note that the equivalence of the more general approaches described in Remark~\ref{remark:general_e} and Remark~\ref{remark:general_A} can be shown by doing the same as above for each $\mathbb{P}\in H_0$.
\end{remark}

\subsection{Multi-stage adaptive designs and e-values\label{sec:multi-stage}}

Several different approaches have been suggested for multi-stage adaptive designs, like multi-stage combination tests, e.g. with inverse normal combination functions (see e.g. \citet{wassmer2016group}), or the recursive combination test \citep{brannath2002recot}. Another, approach is to start with a group sequential test and turn it to an adaptive designs via the conditional rejection probability principle of \citet{muller2004general} (see also Example~\ref{example:mueller}).    
We present here a unified approach, based on a sequence of conditional error functions, which has been introduced implicitly in \citet{brannath2002recot} as alternative representation of recursive combination tests. This extends the above adaptive two-stage designs to multi-stage designs with a possibly infinite and data-driven number of stages. 
%WB: vielleicht auch nicht
We will briefly review recursive combination tests and their equivalence to the here presented adaptive multi-stage designs in Section~\ref{sec:combination_tests}. 

The multi-stage adaptive design consists of a sequence of conditional error functions $A_1, A_2, \ldots$ where each $0\le A_t\le 1$ is a function of the data $D_t$ accumulated until stage $t$ with the property that for all $\mathbb{P}\in H_0$, 
\begin{align}\label{eq:gamd}
\EE_\mathbb{P}[A_1]\le\alpha\quad\text{and}\quad \EE_\mathbb{P}[A_{t}|A_1,\ldots,A_{t-1}]\le A_{t-1} \text{ for all } t=2,3, \ldots .     
\end{align}
With this approach, we can stop and reject $H_0$ at stage $t$ if $A_t=1$. Moreover, we would need to stop at $t$ with an acceptance of $H_0$ if $A_t=0$. Control of the type I error rate at level $\alpha$ with this approach follows immediately from Ville's inequality or the optional stopping theorem, since $(A_t)_{t\in \mathbb{N}}$ is a nonnegative supermartingale with $\EE[A_1]\le\alpha$ under $H_0$. If we decide (at stage $t-1$) to stop the trial with stage $t$, we ideally would choose a binary conditional error function $A_t\in\{0,1\}$, which corresponds to a test decision function that must have a conditional type I error rate of at most $A_{t-1}$.

In order to safely meet condition \eqref{eq:gamd}, the design for the data recruited between stage $t$ and $t+1$, as well as the conditional error function $A_{t+1}$ as function of the data $D_{t+1}$ available at stage $t+1$, need to be specified at stage $t$. Accordingly, the design of the first stage data and the conditional error function $A_1$ need to be specified before the trial.

% Given these conditional error functions, we reject $H_0$ at stage $k$ if $A_k=1$, accept $H_0$ if $A_k=0$ and otherwise continue to stage $k+1$ for $k<K$. In addition, we assume $A_k=1$ whenever $A_{k-1}=1$ to avoid conservative tests. Since $A_K=1$ whenever we reject $H_0$ at any stage, type I error rate control is achieved if $\EE[A_K]\le \alpha$ which follows from \eqref{eq:gamd} and the tower property of conditional expectations. 

\begin{remark}\label{remark:multi-stage}
    In the same manner as in Remark~\ref{remark:general_A}, one could generalize this by choosing a different $A_t^{\mathbb{P}}$ for each $\mathbb{P}\in H_0$, rejecting $H_0$
    at stage $t$ if $A_t^{\mathbb{P}}=1$ for all $\mathbb{P}\in H_0$ and accepting $H_0$ if $A_t^{\mathbb{P}}=0$ for at least one $\mathbb{P}\in H_0$.
\end{remark}

% The general approach mentioned in Remark~\ref{remark:multi-stage} is in line with the conditional error rate principle suggested in \citep{muller2004general} where $A_t^{\mathbb{P}}=\EE_\mathbb{P}[\psi|D_{t}]$ for some prespecified test decision $\psi$ e.g.\ from a group sequential design, where $D_t$ denotes that data at time $t$. We note that the above general approach for multi-stage adaptive designs covers all of the before mentioned (and many more) adaptive multi-stage designs. In particular, it does not require the prespecification of an initial test decision function $\psi$ like for the conditional rejection probability principle \cite{muller2004general}.    

%On the other hand, given sequential e-values $E_1,E_2, \ldots$, we can define $A_t = 1/(\alpha M_t)$ and obtain property \eqref{eq:gamd} from \eqref{eq:evprop}. Again the required restriction on %the support of $A_t$ can be achieved by suitable choice of the e-values. In this case, $A_t=1$ if and only if $M_t=1/\alpha$.

\paragraph{Equivalence to any-time valid inference with sequential e-values.}

It is not difficult to see that our multi-stage adaptive designs can be reformulated in terms of the sequential e-values $E_1=A_1/\alpha$ and $E_t=A_t/A_{t-1}$ that imply $M_t=A_t/\alpha$ for all $t$, where $M_t=\prod_{i=1}^tE_i$. Obviously, \eqref{eq:gamd} implies \eqref{eq:evprop} and $M_t\ge 1/\alpha$ is equivalent to $A_t=1$. 

Conversely, every anytime-valid test based on e-values can be understood as multi-stage adaptive design. To see this, assume given sequential e-values $E_1,E_2,\ldots$ satisfying \eqref{eq:evprop} and $M_t=\prod_{s=1}^t E_s$. Apparently, the anytime-valid test rejects $H_0$ at stage $t$ whenever $\alpha M_t\ge 1$. 
Note that, for a given level $\alpha$, we get an equivalent sequential, anytime-valid test if we use the truncated process $A_t:=\min(\alpha M_t,1)$ and reject $H_0$ stage $t$ if $A_t=1$, simply because $A_t=1$ is equivalent to $\alpha M_t\ge 1$. Obviously, 
$\EE_{\mathbb{P}}[A_1]\le \alpha \EE_{\mathbb{P}}[M_1]\le \alpha$, and for all 
$\mathbb{P}\in H_0$:
$$\EE_{\mathbb{P}}[A_t|E_1,\ldots,E_{t-1}]\le \min(\alpha \EE_{\mathbb{P}}[M_t|E_1,\ldots,E_{t-1}],1)\le \min(\alpha M_{t-1},1)=A_{t-1}.$$ 
This implies that $(A_t)_{t\in\mathbb{N}}$ satisfies \eqref{eq:gamd}, which means that the anytime-valid level-$\alpha$ test based on $(M_t)_{t\in\mathbb{N}}$ is equivalent to the adaptive multi-stage test with conditional error functions $A_t:=\min(\alpha M_t,1)$.

\section{Combination and recursive combinations tests\label{sec:combination_tests}}

Combination tests are an approach to construct adaptive designs, or equivalently anytime-valid tests, directly using p-values. 
It has been shown by \citet{poschbauer1999} and \citet{wassmer1999habil}
that every adaptive two stage design that is based on stage-wise p-values can be written as combination test; see also Wassmer and Brannath (2006).
%or via a conditional error function, the latter being a  function of the first stage p-value  
This statement has been extended in \citet{brannath2002recot} to multi-stage adaptive designs based on so called recursive combination tests which generalize two-stage combination tests to rather flexible multi-stage designs. 
 Recursive combination tests also come with an overall p-value and valid confidence intervals \citep{brannath2002recot}. 
In this section we briefly review combination and recursive combination tests. 
% and show that when a least favorable null distribution for the sequential e-values can be identified, the corresponding anytime-valid test can be rewritten as recursive combination test, and can be uniformly improved if all least favorable (conditional) null distributions are continuous. This is achieved by exhausting the (conditional) levels with improved critical values of the initially equivalent recursive combination test.

\subsection{Two-stage combination tests}\label{sec:combtests}

As indicated in Example~\ref{example:fisher}, combination tests are defined in terms of stage-wise p-values. This means that for a two-stage combination test, the first stage data is summarized by a conservative p-value, i.e.\ a statistic $P_1$ that satisfies 
  \begin{align}\label{eq:cons} 
   \mathbb{P}[P_1\le u]\le u\text{ for all }u\in[ 0,1]\text{ and }\mathbb{P}\in H_0\,.
  \end{align} 
The design and test procedure underlying $P_1$ must be prespecified before the trial.
In addition, there can be preplanned early rejection and acceptance boundaries $\alpha_1<\alpha<\alpha_0$ and, at the first stage, the null hypothesis is rejected if $P_1\le \alpha_1$ and accepted if $P_1>\alpha_0$. For the second stage, a combination function 
  $C:(\alpha_1,\alpha_0]\times [0,1]\to \mathbb{R}$ need to be predefined that is non-decreasing in the first and
increasing in the second argument, and continuous in both arguments. Finally, a critical value $c$ is determined such that  
\begin{align}\label{eq:ct_lc}
\alpha_1+\int_0^1\int_{\alpha_1}^{\alpha_0} \mathbf{1}_{\big\{C(u,v)\le c\big\}}\,du\, dv=\alpha\,.
\end{align}
If the study is continued to the second stage, then the design (e.g.\ sample size) and test procedure for the second stage must be specified at the end of the first stage. For this we can utilize all the information gathered so far. Based on the chosen second stage design and test, a second stage p-value $P_2$ is calculated which must satisfy for all $\mathbb{P}\in H_0$,
\begin{align}\label{eq:pclud}
\mathbb{P}[P_2\le u|P_1]\le u \text{ for all $u\in [0,1]$ and $P_1$.} 
\end{align}
This property is denoted as `p-clud' property in \citet{brannath2002recot}, abbreviating that the \textbf{p-}value $P_2$ is, 
\textbf{c}onditionally on $P_1$, stochastically \textbf{l}arger than or equal to the \textbf{u}niform \textbf{d}istribution. Rejecting $H_0$ if  $C(P_1,P_2)\le c$ then provides a level-$\alpha$ test. Moreover, if there exist a least favorable $\mathbb{P}_0\in H_0$ for which $\mathbb{P}_0[P_1\le u]=u$ and 
$\mathbb{P}_0[P_2\le u|P_1]=u$ for all $u\in [0,1]$ and $P_1$, then the combination test exhausts the level $\alpha$.

A combination test comes with the conditional error function $A(P_1)$ which  equals $1$ for $P_1\le \alpha_1$ and $0$ for $P_1>\alpha_0$, and otherwise is the decreasing function 
$A(P_1):=\sup\{v\in [0,1]:C(P_1,v)\le c\}$ for $P_1\in (\alpha_1,\alpha_0]$. From this, together with the conservatism of $P_1$, we obtain for all $\mathbb{P}\in H_0$ that
$$\mathbb{E}_\mathbb{P}[A(P_1)]\le \alpha_1+\int_{\alpha_1}^{\alpha_0} A(u) du=\alpha\,.$$

Furthermore, a combination test directly provides an overall p-value $Q=Q(P_1,P_2)$, i.e.\ a $[0,1]$-valued random variable with the following properties: 
\begin{enumerate}
\item[(1)] When the stage-wise p-values satisfy \eqref{eq:cons} and \eqref{eq:pclud},  
then $$\mathbb{P}[Q\le u]\le u\text{ for all $u\in [0,1]$ and $\mathbb{P}\in H_0$.}$$ 
Moreover, $Q$ is uniformly distributed on $[0,1]$ for some $\mathbb{P}\in H_0$, if \eqref{eq:cons} and \eqref{eq:pclud} hold with equality for all $u\in [0,1]$ for the same
$\mathbb{P}\in H_0$.
\item[(2)] $Q=P_1$ if the combination test stops at stage 1.
\item[(3)] $Q\le \alpha$ if and only if the combination test rejects $H_0$.
\end{enumerate}
Note that (2) implies $Q\in [0,\alpha_1]\cup (\alpha_0,1]$ if the combination test stops at stage 1 and $Q\in (\alpha_1,\alpha_0]$ if it passes to the second stage. This implies that we can read of from $Q$, whether the trial has stopped at stage 1 or continued to stage 2. This also implies, that when stopping at stage 1, $Q\le \alpha$ and $Q> \alpha$ are equivalent to $P_1\le \alpha_1$ and  $P_1>\alpha_0$, respectively. 
The construction of the p-value $Q$ depends only on $\alpha_1$, $\alpha_0$ and $C$; see \citet{brannath2002recot}.

\subsection{Recursive combination tests}  

We sketch now the idea of recursive combination tests. A detailed description and exploration can be found in \citet{brannath2002recot}. 

A recursive combination test starts with a prespecified two-stage combination test, 
i.e. a prespecified first stage design, corresponding first stage p-value $P_1$, first stage rejection and acceptance levels $\alpha_{1,1}<\alpha<\alpha_{1,0}$, and continuous and monotone combination function $C_1:(\alpha_{1,1},\alpha_{1,0}]\times [0,1]\to \mathbb{R}$.
Like for two-stage combination tests, the first argument of the combination function $C_1$ is for the first stage p-value $P_1$. However, the second argument of $C_1$ is a now a placeholder for the evidence that will come from all, yet unspecified forthcoming stages. If this evidence is summarized by a p-value $Q_2$ with the p-clud property, i.e.\ $\mathbb{P}[\,Q_2\le u\,|\,P_1\,]\le u$
for all $u\in [0,1]$ and $\mathbb{P}\in H_0$, then we obtain a level-$\alpha$ test and valid p-value $Q_1$, when using $Q_2$ as second argument of $C_1$. 

With a recursive combination test, we use the p-value $Q_2$ from another combination test, now at level $A_1(P_1)$, the conditional error function of the first combination test. The procedure can be repeated by using for the second stage p-value of the second combination test the p-value function of another combination test, now at the level that equals the conditional error function of the second combination test, etc. This procedure can be repeated as long as we wish. If we decide at some stage $t$ to finish the study with the next stage $t+1$ (and have not stopped the study yet with an early rejection or acceptance of $H_0$), then we use as second argument of the final combination test a final (conditional) p-value $P_{t+1}$. 
At each stage of this procedure we can decide based on all internal and external information obtained so far, whether to stop the study with the next stage or continue with at least two further stages. Thereby we can choose at each stage, based an all information gathered so far, the design and test procedure for the next stage and, in case we decide to continue beyond the next stage, the combination test for merging the next and forthcoming stages.   

It has been shown in \citet{brannath2002recot} that this recursive procedure controls the type I error rate whenever we use at each stage $t$ a p-value $P_t$ for the first argument of the combination test that satisfies 
the sequential p-clud condition
\begin{align}\label{eq:spclud}
   \mathbb{P}[P_t\le u|P_1\ldots,P_{t-1}]\le u\text{ for all $u\in [0,1]$ and } \mathbb{P}\in H_0\text{ almost surely.}
\end{align}

Moreover, the conditional error functions $A_t$ of the recursively defined combination tests 
satisfy \eqref{eq:gamd} under the sequential p-clud condition \eqref{eq:spclud}. In addition, we obtain 
$\mathbb{E}_{\mathbb{P}}[A_1]=\alpha$ and 
$\mathbb{E}_{\mathbb{P}}[A_t|A_1,\ldots,A_{t-1}]=A_{t-1}$ almost surely for all $t$, whenever the p-values $P_t$
are independent and uniformly distributed on $[0,1]$ under $\mathbb{P}$, i.e. 
satisfy \eqref{eq:spclud} with equality for all $u\in [0,1]$.

Conversely, a sequence of conditional error functions $A_t$, $t=1,2,\ldots$, with the properties given in Section~\ref{sec:multi-stage}, can be represented as recursive combination test, if each $A_t=A_t(P_t)$ is a function of a p-value $P_t$ satisfying \eqref{eq:spclud}. The combination functions of this recursive combination test are e.g.\ 
$C_t(P_t,v):=v/A_t$ with critical value $c_t=1$ . Obviously, $A_t=\sup\{v\in [0,1]:C_t(P_t,v)=v/A_t\le 1\}$. 
Note that $A_t$ can be written as function of a p-clud p-value $P_t$, if $A_t$ has a known least favorable conditional 
survival function $S_t(u)$ under the null hypothesis, i.e.\ $S_t(u)\ge \bbP[A_t\ge u| A_1,\ldots, A_{t-1}]$ for all $\bbP\in H_0$ and $u\in [0,1]$ almost surely, namely $A_t=S_t^{-1}(P_t)$ with $P_t=S_t(A_t)$, where $S_t^{-1}$ is the (generalized) inverse of $S$.
%By this, we obtain that also any-time valid tests based on sequential e-values are equivalent to recursive combination tests. 

We show in Appendix~\ref{sec:exhaust} that (and how) sequential e-values can be uniformly improved by recursive combination tests, if they do not exhaust the level and have known and continuous least favorable conditional null distributions. 

%Finally, we can calculate an overall p-value $Q$ for the recursive combination test
%utilizing the p-values $q_t$ of the (conditional) combination tests with the following %recursive backward algorithm whereby we assume that $t+1$ is the last stage: 
%$$Q_{t+1}=q_{t}(P_t,P_{t+1})\text{ and } Q_s=q_{s-1}(P_s,Q_{s+1})\text{ for } s=t,\ldots,1.$%$ 
%One can verify by backward induction in $s$, that $Q$ satisfies $\mathbb{P}[Q\le u]\le u$
%for all $u\in [0,1]$ and $\mathbb{P}\in H_0$ under the sequential p-clud property %\eqref{eq:spclud}, and if the sequential p-values $P_t$
%are stochastically independent and uniformly distributed 
%then, $Q$ is uniformly distributed on $[0,1]$ as well.

%%%%%%%%%%%%%%%%%%%%%%%%%%%%%%

\section{Multiple testing in adaptive designs and e-values\label{sec:multiple_testing}}   
Adaptive designs are particularly useful when multiple hypotheses are tested, since this, for example, allows to drop less efficient treatment arms at an interim analysis and/or add new more promising ones. In clinical trials, it is often required to control the familywise error rate (FWER), and in adaptive designs, this can be achieved by an application of the closed testing principle with the above reviewed adaptive testing approaches, e.g.\ with conditional error functions or combination tests, see e.g. \cite{Kieser1999combining,hommel2001modif,posch2011type,wassmer2016group}. A popular multiple testing procedure for clinical trials with multiple hypothesis is the graphical approach by \citet{bretz2009graphical}, which is based on the (more general) consonant weighted Bonferroni closed tests considered in \cite{hommel2007short-cuts} and has been extended to adaptive designs in  
\cite{klinglmueller2014adaptive}.
In the following, we connect the extension to adaptive designs to formally similar suggestions from the e-value literature.
%It should be noted that the presented methods, although conceptually similar as the graphical procedure \citep{bretz2009graphical}, cannot %necessarily represented by a graph.

%The graphical procedure \citep{bretz2009graphical} is constructed using the closure principle \citep{marcus1976closed} with weighted Bonferroni tests for each intersection hypothesis. That means, 
We start reviewing the mentioned closed test procedure. Given p-values $P_1,\ldots, P_m$ for $m$ null hypotheses $H_0^1,\ldots,H_0^m$ of interest, we reject the intersection hypothesis $H_0^I=\bigcap_{i\in I} H_0^i$, for $I\subseteq\{1,\ldots,m\}$, if $\min_{i\in I} P_i \leq w_i^I \alpha$, where the $(w_i^I)_{i\in I}$ are prespecified weights such that $\sum_{i=1} w_i^I=1$ and $w_i^I\geq w_i^J$ if $i\in J\subseteq I$. The closed procedure
%which results in the graphical procedure in this case, 
then rejects an individual hypothesis $H_0^i$, $i\in \{1,\ldots,m\}$, if all $H_0^I$ with $i\in I$ are rejected. 

One can naturally extend this procedure to adaptive designs by applying the conditional rejection probability principle (see Example~\ref{example:mueller}) to each of the marginal test decision functions $\varphi^I_i=\mathbf{1}_{\{P_i \leq w_i^I \alpha\}}$ for $i\in I\subseteq \{1,\ldots,m\}$. This leads to the in \cite{posch2008Bonferroni} introduced partial conditional error functions $A_i^I=\mathbb{E}_{\mathbb{P}}[\varphi^I_i|D]$.
Here, $D$ represents the interim data; see Example~2 of this paper's introduction. A nice property of the partial conditional error functions is, that for each $\mathbb{P}\in H^I_0$, the their sum satisfies 
$\mathbb{E}_{\mathbb{P}}[\sum_{i\in I}A^I_i]=\sum_{i\in I}\mathbb{E}_{\mathbb{P}}[\varphi^I_i]\le \sum_{i\in I} w_i^I \alpha=\alpha$. Therefore, $A_I=\min\left(\sum_{i\in I}A^I_i,1\right)$ provides a conditional error function for $H^I_0$ that permits (in addition to data driven sample size adaptations) to drop hypotheses as well as changing the weights for the remaining ones; see \cite{klinglmueller2014adaptive} for details. 
% A major advantage of this method is that only requires the knowledge of
% the marginal distributions of the p-values or the underlying test statistics. 
Moreover, if $A_I=1$, which is possible even when every $A^I_i<1$, then one can reject $H_0^I$ already at the interim analysis. The latter leads to a costless uniform improvement of the single stage weighted Bonferroni closed test for $H_0^I$; see also \cite{posch2008Bonferroni}.

We note that this approach is possible with any (marginal) conditional error function $A_i^I$ at level $\alpha w_i^I$, meaning that
\begin{align}\mathbb{E}_\mathbb{P}[A_i^I]\leq \alpha w_i^I \quad \text{for all distributions } \mathbb{P}\in H_0^i.
\label{eq:mult_test_cond_err} \end{align} 
We should also note, that when the joint null distribution of the p-values can be utilized, alternative and eventually more efficient adaptive closed test procedures are available; see e.g.\ \cite{wassmer2016group}. 

%Due to the linearity of the expectation, one obtains a conditional error function for $H_0^I$ at level $\alpha$ by $A_I=\min\left(\sum_{i\in I} A_i^I,1\right)$. If $A_i^I= 1$ for at least one $i\in I$, then $A_I=1$ and we can reject $H_0^I$ at the first stage. If $A_i^I=0$ for all $i\in I$, then $A_I=0$ and need to accept $H_0^I$ at stage 1. Note that $A_I=1$ is possible also if all $0<A_i^I<1$, which would permit to reject $H_0^I$ already at stage 1.
%If $0<A_I<1$, we can decide to continue with the second stage and fix its design together with a second stage conditional error (or test decision) function for each $H_0^I$ whose conditional expectation (given the first stage data) is bounded by $A_I$ under all null distributions. 
%collect more data and calculate another conditional error function for $H_0^I$ in the same manner as above. 
%This proceeding, although slightly different formulated, was proposed by \citet{klinglmueller2014adaptive}. 

Equivalently to the adaptive closed test based on partial or marginal conditional error funcions, we could define an e-value $E_i^I$ for $H_0^i$ and each $I\subseteq \{1,\ldots,m\}$ by 
$E_i^I={A_i^I}/{\alpha w_i^I}$, 
where the e-value property follows by \eqref{eq:mult_test_cond_err}. An e-value $E^I$ for $H_0^I$ can then be constructed by taking the weighted average $E^I=\sum_{i\in I} w_i^I E_i^I$, a common strategy to merge arbitrary e-values into a single e-value \citep{vovk2021values}. 

Since $E^I\geq 1/\alpha$ is equivalent to $A^I=1$, the closed procedure based on the e-values $(E^I)_{I\subseteq \{1,\ldots,m\}}$ rejects the same hypotheses as the closed procedure based on the conditional error functions $(A^I)_{I\subseteq \{1,\ldots,m\}}$. \citet{hartog2025family} considered the case where the same e-value $E_i=E_i^I$ is used for each intersection hypothesis $H_0^I$ with $i\in I$ and provided computationally efficient short-cuts for particular weight combinations $(w_i^I)_{i\in I}$.

Remarkably, the strategy to merge e-values by a weighted average is the only admissible one in the sense that, any other strategy to combine several e-values to a single e-value, that works without any further distributional assumptions, can be improved by a weighted average of the e-values; see \citep{wang2025only}. By the equivalence between e-values and conditional error functions, this result also applies to the combination of conditional error functions.

\section{Discussion\label{sec:discussion}}
In this note, we have linked the currently disjoint literature on adaptive designs and e-values. In particular, we showed that both approaches allow to construct the same sequential tests and are closely related from a methodological perspective. We hope this will make the literature more accessible to people from each other's communities and eliminate confusion for those new to sequential and adaptive hypothesis testing.

Please be aware that this note is not intended to downplay the new achievements of the e-value literature. On the contrary, we are convinced that the e-value perspective can have many advantages and has already led to several groundbreaking results. In the following, we summarize some of these results and briefly discuss the extent to which they may also be relevant for the design of clinical trials. 

Afterwards, we note some advantages of the classical approaches to construct adaptive designs.  This highlights the practical value of this paper, as it allows researchers to switch between the different adaptive approaches and thus take advantage of all of the methods.

\subsection{Advantages of e-values}
\paragraph{Post-hoc choice of loss and significance level.}
\citet{grunwald2024beyond} introduced a testing framework in which type I error control is replaced by expected loss control. While this gives a reformulation of the classical Neyman-Pearson setup for a fixed loss function, \citet{grunwald2024beyond} goes much beyond this by allowing the loss function to depend on the data in an arbitrary way. He shows that the use of e-values retain their guarantee for such data-dependent losses, while p-values do not, and that all admissible tests with such post-hoc guarantee must be based on e-values. This is comparable to type I error control at a data-dependent level $\alpha$ (in a specific sense; see \citet{koning2023post} for more insights). Note that such a guarantee is not possible with conditional error functions, since the $\alpha$ must be prespecified there.

We believe that in clinical trials the risk of committing a type I error often varies due to external factors as well as the observed data such that the aforementioned framework captures situations that are difficult or even impossible to handle with the currently available statistical methodology for clinical trials. For example, consider two treatments  that are tested for the same disease. Suppose the first treatment showed no safety issues (compared to the control), while adverse events were observed more frequently for the second treatment. It would be natural to set the bar for approving the second treatment, meaning the loss of a type I error regarding the efficacy, higher than for the first treatment. There are many further clinical trial scenarios in which the loss of a type I error cannot be chosen upfront \citep{grunwald2024beyond}.

% \citet{grunwald2024beyond} introduced a testing framework in which discrepancies between the test decision $\phi$ and the true state of nature $\kappa$ ($\kappa=0$ if $H_0$ is true and $\kappa=1$ otherwise) are described by a loss function $L(\kappa, \phi)$. The classical type I control $\mathbb{P}_{H_0}(\phi=1)\leq \alpha$ is replaced by requiring that the expected loss of falsely rejecting the null hypothesis is bounded by one:
% \begin{align}
% \mathbb{E}_{H_0}[L(0,\phi)]\leq 1. \label{eq:loss_simple}
% \end{align}
% If $L(0,1)=1/\alpha$, this is equivalent to the classical type I error control. However, \citet{grunwald2024beyond} goes much beyond this by allowing multiple loss functions $L_b(\kappa, \phi)$, $b\in \mathcal{B}$, (and also multiple actions instead of binary decisions, which we skip here for brevity) where the set $\mathcal{B}$ indexes the loss functions.

% \citet{grunwald2024beyond} shows that the e-value based test that rejects $H_0$ with a loss of $L_b(0,1)$ in case of $E\geq L_b(0,1)$, $\phi_b\coloneqq \mathbbm{1}\{E\geq L_b(0,1)\}$, where $E$ is some e-values, guarantees that 
% $$
% \mathbb{E}_{H_0}\left[\sup_{b\in \mathcal{B}} L_b(0,\phi_b) \right]\leq 1,
% $$
% which allows to choose the a specific loss function $L_b(0,1)$ based on the data while ensuring that \eqref{eq:loss_simple} is fulfilled. 

\paragraph{Log-optimality.}
In contrast to fixed sample size tests, where a uniformly most powerful test often exists, there are usually many possible ways to  construct a sequential test, and thus many possible e-values and conditional error functions, due to different weightings of the stages. However, the e-value literature has developed a well-founded choice \citep{shafer2021testing, grunwald2020safe, ramdas2024hypothesis}: the \textit{log-optimal} e-value maximizes $\mathbb{E}_{\mathbb{Q}}[\log(E)]$ under some alternative distribution $\mathbb{Q}$ of interest.  

The log-optimality criterion dates back to \citet{kelly1956new} and \citet{breiman1961optimal} who studied optimal gambling strategies. The idea is that if many i.i.d. e-values are multiplied, then the law of large number applies to the logarithm of the product, and thus log-optimal factors guarantee the fastest possible growth rate. There is already very sophisticated theory on the log-optimal e-value, showing its existence under no assumptions about the null hypothesis and a characterization closely related to likelihood ratios \citep{grunwald2020safe, larsson2024numeraire}. This may provide general guidance on how to construct powerful adaptive designs. However, it should be noted that the optimality of the criterion is built on the premise of a potentially infinite number of stages. If practical constraints limit the number of stages to just a few, other criteria may be more appropriate.

\paragraph{Multiple testing.}
Recently, \citet{xu2025bringing} introduced a closure principle that is based on local e-values for the intersection hypotheses (instead of local tests) and which can be used to construct powerful procedures for any expectation based error metric, including, e.g.,  FWER, false discovery rate (FDR) \citep{benjamini1995controlling} and per-family error rate (PFER) \citep{benjamini1997multiple}. This general approach also applies to weighted variations of these error rates, which have been recently discussed for settings where treatments are tested in multiple subpopulations \citep{maurer2023optimal, brannath2023population}. Similar to the post-hoc choice of the loss function in the testing framework of \citet{grunwald2024beyond}, this e-value based closure principle allows to choose the error metric and the significance level based on the data, providing enormous flexibility that has potential for exploratory analyses.

\paragraph{Universal inference.}
\citet{wasserman2020universal} introduced a general approach to hypothesis testing that works in irregular statistical models, where the null distribution of the classical likelihood ratio statistic is often intractable. While p-values cannot be directly computed in such complex testing problems, the \textit{universal inference} approach by \citet{wasserman2020universal} uses e-values by construction. 
Universal inference also offers a simple sequential version that can be useful in complex testing problems for which adaptive designs have not been constructed yet.

\paragraph{Interpretation.} 
\citet{shafer2021testing} introduced a testing by betting approach, where e-values can be interpreted as the betting score in a fair game against the null hypothesis. This, as well as its error control at data-dependent significance levels \citep{grunwald2024beyond, koning2023post}, allow e-values to be interpreted as \textit{continuous measure of evidence}. Regardless of the chosen significance level $\alpha$, a large e-value provides evidence against the null hypothesis. Note that this is not true for conditional error functions, as their interpretation is strongly tied to the prespecified significance level. 

In particular, this makes it sensible to report an e-value as result of a study. Further, due to the optional stopping theorem, one could then multiply e-values of future experiments with that e-value
to combine the evidence, providing a strong and valid method for meta-analyses. Moreover, testing by betting \citep{shafer2021testing} provides an intuitive approach to hypothesis testing that prevents misuse and can help with teaching and communication of adaptive designs to students and applicants.

\subsection{Advantages of adaptive designs}

\paragraph{Conditional error function principle.}
Conditional error functions give the level that can be used for future tests and therefore provide an intuitive interpretation of the adaptive design. In particular, this is useful when conditional sample size adjustments or power calculations are performed during interim analyses, as those typically require a specified significance level. Furthermore, conditional error functions can directly be obtained by the conditional rejection principle (see Example~\ref{example:mueller}) and exhaust the type I error if \eqref{eq:conderr1} and \eqref{eq:conderr2} are satisfied with an equality.

\paragraph{(Recursive) combination tests.}
An advantage of combination tests are that they directly work with p-values. As there are standard p-values for many different testing problems, combination tests are often easy to apply. In addition, combination tests provide an overall p-value that can be used to summarize the result of an adaptive design. Furthermore, they exhaust the type I error level by design as long as the p-values are exact (uniformly distributed under the null hypothesis). This is not the case for e-value based anytime-valid tests, which can often be uniformly improved exploiting their representation as (recursive) combination test; see Appendix~\ref{sec:exhaust} for more details. 

\subsection*{Acknowledgments}
 LF acknowledges funding by the Deutsche Forschungsgemeinschaft (DFG, German Research Foundation) – Project number 281474342/GRK2224/2. 

\bibliography{main}
\bibliographystyle{plainnat}

\begin{appendix}

\section{Exhausting e-value based tests with recursive combination tests}\label{sec:exhaust}
It is well known that $1/U$ is not an e-value for a uniformly distributed $U$. This is because $1/U$ has infinite expectation. By this, 
the reciprocal $1/E_1$ of an e-value $E_1$ must be a strictly conservative p-value (when truncated at $1$),  i.e.\ $\mathbb{P}(1/E_1\le u)\le u$ for all $u\in[0,1]$ and $\mathbb{P}\in H_0$ (by the Markov inequality), with a strict inequality for at least one $u$. If $E_1$ is continuously distributed  under $\mathbb{P}$, then the strict inequality holds on a whole interval of $[0,1]$. 
If $\mathbb{P}_0$ is a known least favorable configuration, i.e. $S_1(u):=\mathbb{P}_0(1/E_1\le u)\le \mathbb{P}(1/E_1\le u)$ for all $u\in [0,1]$ and $\mathbb{P}\in H_0$,
then we can define the p-value $P_1=S_1(1/E_1)$, which is almost surely smaller than or equal to $1/E_1$, with a positive probability for $P_1<1/E_1$, if $E_1$ is continuously distributed under $\mathbb{P}_0$. 

For the sequential e-values $E_t$, $t\ge 2$, the same is true conditionally. In detail, let $$S_t(u):=\mathbb{P}_0(1/E_t\le u|E_1,\ldots,E_{t-1}),$$ 
where $\mathbb{P}_0\in H_0$ is a known least favorable configuration for all $E_t$, i.e. $S_t(u)\le \mathbb{P}(1/E_t\le u|E_1,\ldots,E_{t-1})$ for all $u\in [0,1]$ and $\mathbb{P}\in H_0$.
This permits us to define the 
sequence of p-values $P_t=S_t(1/E_t)$ that satisfies the sequential p-clud property \eqref{eq:spclud}, and $\mathbb{P}_0[P_t<1/E_t|E_1,\ldots, E_{t-1}]>0$ almost surely when $u\mapsto S_t(u)$ is continuous  (which often applies to likelihood ratios). 
This indicates that we may be able to uniformly improve the given e-value test by a recursive combination test.  We indicate below how this can done, providing the details in the appendix. 

We start with the combination test that has first stage rejection boundary $\alpha_{1,1}:=S_1(\alpha)$ and the combination function $C_1(P_1,P_2)=P_2/E_1$ with critical value $c_1=\alpha$, whereby $E_1=1/S_1^{-1}(P_1)$ can be understood as a function of the first stage p-value $P_1=S_1(1/E_1)$. When stopping the study at the second stage and using $1/E_2$ for the second stage p-value $P_2$, this combination test is equivalent to the test based on the e-values $E_1$ and $E_2$ which rejects $H_0$ if $E_1 E_2\ge 1/\alpha$. This two stage combination test has conditional error function $A_1(P_1)=\sup\{u\in [0,1]:C_1(P_1,u)=u/E_1\le \alpha\}=\min(\alpha E_1,1)$ which is identical to the first stage conditional error function of the sequential e-value test; see Subsection~\ref{sec:multi-stage}. 

Moreover, when $S_1(\alpha)=\mathbb{P}_0[\alpha E_1\ge 1]>0$, i.e.\ we have the chance to reject $H_0$ with $E_1$ at stage 1, then we get
\begin{align}\label{eq:conservative}
\mathbb{E}_0[\min(\alpha E_1,1)]=S_1(\alpha)+\mathbb{E}_0[\min(\alpha E_1,1)\mathbf{1}_{\{P_1> S_1(\alpha)\}}]
<\alpha
\end{align}
where $\mathbb{E}_0$ is the expectation under $\mathbb{P}_0$. This follows from 
$$\alpha\ge \mathbb{E}_0[\alpha E_1]= \mathbb{E}_0[\min(\alpha E_1,1)]+\mathbb{E}_0[(\alpha E_1-1)\mathbf{1}_{\{\alpha E_1>1\}}]$$
and $
\mathbb{E}_0[(\alpha E_1-1)\mathbf{1}_{\{\alpha E_1>1\}}]>0$. The latter can be seen  by the fact that for a continuously distributed $E_1$:
$\mathbb{P}_0[\alpha E_1> 1]=\mathbb{P}_0[\alpha E_1\ge 1]>0$ implies $\mathbb{P}_0[\alpha E_1-1\ge \epsilon]>0$ for some $\epsilon>0$. 
Note that $\mathbb{P}_0[\alpha E_1-1\ge \epsilon]>0$ for some $\epsilon>0$ is also satisfied for a non-continuously distributed $E_1$,
whenever there exists some $\alpha'<\alpha$ such that $\mathbb{P}_0[E_1\ge 1/\alpha']>0$, i.e.\ $E_1$ is not bounded by $1/\alpha$.

The left side of inequality in \eqref{eq:conservative} equals the level \eqref{eq:ct_lc} of the combination test. Hence, the equivalent combination test is conservative even with independent and uniformly distributed stage-wise p-values and can be uniformly improved by increasing the first stage level $\alpha_{1,1}$ and/or $c_1$ such that its level $\alpha$ is exhausted. When the trial is stopped at the second stage, the use of $P_2=S_2(1/E_2)$ will often uniformly improve the e-value based test further.

Let us turn now to the case when the sequential e-value test continuous beyond stage 2. One can easily see that the e-value based test at stage $t\ge 2$ is equivalent to the combination test with early rejection level $\alpha_{t,1}=S_t(\alpha M_{t-1})$ and the combination function $C_t(P_t,u)=S_t^{-1}(P_t)\cdot u$ with critical value
$c_t=\alpha M_{t-1}$. By similar arguments as for the first two stages, the rejection region of this combination test does not exhaust the conditional level of the e-value based test, and even less the larger conditional level $A_{t-1}$ of the combination test from the previous stage that was improved to exhaust the conditional level of stage $t-2$. As before, the conditional level $A_{t-1}$ can be exhausted with the given combination function $C_t$ by increasing $\alpha_{t,1}$ and/or $c_t$, in a way that uniformly improves the equivalent combination test and thereby the sequential e-value based test. %This combination test provides a conditional error function $A_t$
%which can be exploited with the next combination test. 
Proceeding in this way at every stage, we obtain a sequence of conditional error functions that satisfies 
$A_{t-1}=\mathbb{E}_0[A_t|E_1,\ldots,E_{t-1}]$ almost surely at each stage $t$,
and $\mathbb{E}_0[A_1]=\alpha$. 
If $t$ is the last stage, then we finally can exhaust the level $\alpha$ by using $P_t=S_t(1/E_t)$ for the final p-value in the final combination test (instead of the larger p-value $1/E_t$).    

\begin{remark} We show in the Appendix that a sequential test based on two e-values and with a positive chance to reject $H_0$ at the first stage, is inevitably conservative, i.e.\ satisfies \eqref{eq:conservative},
whenever $\mathbb{P}_0[E_1\ge 1/\alpha']>0$ for some $\alpha'>\alpha$. This
holds in general, i.e.\ also for an e-value $E_1$ that is not continuously distributed (as assumed in this section). 
This finding implies that flexibility with respect to significance levels necessarily comes for the price of a strict conservatism.
\end{remark}

% \subsection{Conservatism of e-value based tests} \label{sec:cons_tests}
% We show that for an e-value $E_1$, which is continuously distributed on an interval of $\mathbb{R}_+$, the condition $S_1(\alpha)=
% \mathbb{P}_0[\alpha E_1\ge 1]>0$ implies $\mathbb{E}_0[\min(\alpha E_1,1)]
% %=S_1(\alpha)+\mathbb{E}_0[\min(\alpha E_1,1)\mathbf{1}_{\{p_1> S_1(\alpha)\}}]
% <\alpha$. This follows from
% $$\alpha\ge \mathbb{E}_0[\alpha E_1]= \mathbb{E}_0[\min(\alpha E_1,1)]+\mathbb{E}_0[(\alpha E_1-1)\mathbf{1}_{\{\alpha E_1>1\}}]$$
% and $
% \mathbb{E}_0[(\alpha E_1-1)\mathbf{1}_{\{\alpha E_1>1\}}]>0$. The latter can be seen e.g.\ by the fact that for a continuously distributed $E_1$:
% $\mathbb{P}_0[\alpha E_1> 1]=\mathbb{P}_0[\alpha E_1\ge 1]>0$ implies $\mathbb{P}_0[\alpha E_1-1\ge \epsilon]>0$ for some $\epsilon>0$. 
% Note that $\mathbb{P}_0[\alpha E_1-1\ge \epsilon]>0$ for some $\epsilon>0$ is also satisfied for a non-continuously distributed $E_1$,
% whenever there exists some $\alpha'<\alpha$ such that $\mathbb{P}_0[E_1\ge 1/\alpha']>0$, i.e.\ $E_1$ is not bounded by $1/\alpha$.

\subsection{Combination tests that exhaust the level of sequential e-value tests}\label{sec:exhausting}

We assume in this subsection that all e-values have (conditional) Lebesque-densities that are strictly positive on $\mathbb{R}^+$ or an interval of $\mathbb{R}^+$, an assumption that often applies to likelihood ratios. 
%Often there exists a least favorable distribution $\mathbb{P}_0\in H_0$, such that
%$\max_{\mathbb{P}\in H_0} \mathbb{P}[E\ge 1/u] = \mathbb{P}_0[1/E\le u]=:S(u)$ for all $u\in [0,1]$ (e.g.\ when $H_0$ is simple) and then we have 
%
%We can use similar arguments to improve a sequential test based on sequential e-values $E_1, E_2, \ldots$, using recursive combination tests. To this end, 
We further assume the existence of a known least favorable null distribution $\mathbb{P}_0\in H_0$, such that $S_1(u):=\mathbb{P}_0[1/E_1\le u]=\max_{\mathbb{P}\in H_0}\mathbb{P}[1/E_1\le u]$ and 
$S_t(u):=\mathbb{P}_0[\,1/E_t\le u\,|\,E_1,\ldots, E_{t-1}\,]=\max_{\mathbb{P}\in H_0}\mathbb{P}[\,1/E_t\le u\,|\,E_1,\ldots, E_{t-1}\,]$ almost surely for all $u\in [0,1]$ and $t\ge 2$. This permits us to define the sequence of p-values $P_t=S_t(1/E_t)$, which satisfy the sequential p-clud property \eqref{eq:spclud}, with equality for all $u\in [0,1]$ under $\mathbb{P}_0$. We will also use the inverse functions $S_t^{-1}(x)=\sup\{x\ge 0: S_t(u)\ge x\}$ that satisfy $S^{-1}_t(P_t)=1/E_t$. Finally, we can assume without loss of generality that $\mathbb{E}_0[E_1]=1$ and $\mathbb{E}_0[E_t|E_1,\ldots,E_{t-1}]=1$ almost surely for all $t$, using from now on 
$\mathbb{E}_0$ for $\mathbb{E}_{\mathbb{P}_0}$. (If this conditional expectation is for some $E_t$ strictly smaller 1, then we can divide $E_t$ by its conditional expectation and obtain a uniformly larger and thereby more powerful sequential e-value.) 

We show how the two-stage combination test, that replicates the first two stages of the sequential e-value test, can be uniformly improved to exhaust the level $\alpha$. The same arguments apply to the combination tests for the later stages with $\alpha$ replaced by the current conditional level. We can exhaust the level of the first combination test by increasing $c_1$ to the larger critical value
$$c^{\max}_{1}:=\max\big\{v\in [0,1]: S_1(\alpha)+\mathbb{E}_0[\min(v E_1,1)\mathbf{1}_{\{P_1> S_1(\alpha)\}}]\le \alpha\big\}$$ 
and still control the type I error rate at level $\alpha$ with independent and uniformly distributed p-values. 

This exhausts the level of the combination test 
if $S_1(\alpha)+\mathbb{E}_0[\min(E_1,1)\mathbf{1}_{\{P_1> S_1(\alpha)\}}]\ge  \alpha$. Note that, in this case, the improvement is also achieved by simply using the p-value function $Q_1$ of the combination test, because this automatically results in using the level exhausting critical value $c^{\max}_{1}$.

If $S_1(\alpha)+\mathbb{E}_0[\min(E_1,1)\mathbf{1}_{\{P_1> S_1(\alpha)\}}]<  \alpha$, we can choose $c^{\max}_1=1$ as critical value for the combination function and exhaust the level by increasing the first stage level 
$\alpha_{1,1}$ to the level $\alpha^{\min}_{1,1} S_1(\alpha)$ which satisfies 
\begin{align}\label{eq:aloneone}
\alpha^{\min}_{1,1}+\mathbb{E}_0[\min(E_1,1)\mathbf{1}_{\{P_1> \alpha^{\min}_{1,1}\}}]=\alpha.
\end{align}

It is worth to mention, that this is not the only way to exhaust the level of the combination test while improving the e-value based test. To this end we can choose any $c_1$ between $c^{\min}_1=\sup\{u\in[0,1]:\mathbb{E}_0[\min(\alpha E,1)]\}$ and $c^{\max}_1$, and then determine $\alpha_{1,1}$ such that $\alpha_{1,1}+\mathbb{E}_0[\min(c_1 E_1,1)\mathbf{1}_{\{P_1> \alpha_{1,1}\}}]=\alpha$ 
 is satisfied. 

We finally note that the uniform improvement of Wald's sequential probability ratio test suggested in \cite{fischer2024improving} corresponds to the specific choice  $\alpha_{1,1}=S_1(\tilde{\alpha})$ and $c_1=\tilde{\alpha}$ for the level $\tilde{\alpha}>\alpha$ that satisfies 
$$\mathbb{E}[\min(\tilde{\alpha} E_1,1)]=S_1(\tilde{\alpha})+\mathbb{E}[\min(\tilde{\alpha} E_1,1)\mathbf{1}_{\{P_1> S_1(\tilde{\alpha})\}}]=\alpha.$$

\end{appendix}

\end{document}